# Near-field Hyperspectral Imaging of Resonant Mie Modes in a Dielectric Island

Nicoletta Granchi[1], Michele Montanari[1], Andrea Ristori[1], Mario Khoury[2], Mohammed Bouabdellaui[2], Chiara Barri[3,4], Luca Fagiani[3,4], Massimo Gurioli[1], Monica Bollani[3], Marco Abbarchi[2,*], and Francesca Intonti[1,**]

[1] LENS, University of Florence, Sesto Fiorentino, 50019, Italy

[2] Aix Marseille Univ, Universitè de Toulon, CNRS, IM2NP, Marseille, France

[3] Istituto di Fotonica e Nanotecnologie (IFN)-Consiglio Nazionale delle Ricerche, Laboratory for Nanostructure Epitaxy and Spintronics on Silicon, Via Anzani 42, 22100 Como, Italy

[4] Department of Physics, Politecnico di Milano, Piazza Leonardo Da Vinci 32, 20133 Milano, Italy

* marco.abbarchi@im2np.fr

** intonti@lens.unifi.it

## Abstract

All-dielectric, sub-micrometric particles have been successfully exploited for light management in a plethora of applications at visible and near-infrared frequency. However, the investigation of the intricacies of the Mie resonances at the sub-wavelength scale has been hampered by the limitation of conventional near-field methods. Here we address spatial and spectral mapping of multi-polar modes of a Si island by hyper-spectral imaging. The simultaneous detection of several resonant modes allows to clarify the role of substrate and incidence angle of the impinging light, highlighting spectral splitting of the quadrupolar mode and resulting in different spatial features of the field intensity. We explore theoretically and experimentally such spatial features. Details as small as 200 nm can be detected and are in agreement with simulations based on a Finite Difference Time Domain method. Our results are relevant to near-field imaging of dielectric structures, to the comprehension of the photophysics of resonant Mie structures, to beam steering and to the resonant coupling with light emitters. Our analysis paves the way for a novel approach to control the spatial overlap of a single emitter with localized electric field maxima.

## Introduction

All dielectric sub-wavelength sized Mie resonators have emerged in the last decade as promising building blocks for optoelectronic devices, since they provide the possibility to efficiently redirect and concentrate light with low absorption losses [1]. The optical modes in high-index dielectric nanoparticles originate from the excitation of optically induced displacement currents, and can be both magnetic and electric in nature. The combination of these two kinds of resonant modes and the exploitation of higher order multipolar modes offer opportunities for directional and polarization controlled emission from nanoemitters [2-6].

Moreover, in the field of light-matter interaction, high refractive index dielectric nanoparticles have recently been considered as a possible alternative to metallic nanoparticles for generating localized optical resonances down to nanoscale. In fact, the concentration of light in subwavelength hot spots represents an essential element for surface enhanced fluorescence [7] and for manipulating the generation of nonclassical light by quantum emitters [8].

For these purposes, a complete knowledge of the electric and magnetic nature of the modes, their spectral features and their spatial distribution is mandatory. Finite difference time domain (FDTD) numerical calculations represent a powerful tool to systematically study the resonant properties of high-index dielectric particles [9-11]. However, the direct experimental characterization of these peculiarities has almost exclusively dealt with far-field measurements [12,13], that do not provide access to the spatial distribution of light modes.

Sub-wavelength, spatial resolution imaging has been realized in the past by cathodoluminescence imaging spectroscopy [14] or apertureless (scattering type) near-field scanning optical microscopy [15-17]. However, these methods are either limited to the collection of the signal in dielectric regions, i.e. only in the small area covered by the sub-wavelength sized Mie resonators [14], or do not provide information on more than one optical mode at a time, owing to the use of a laser at a fixed frequency, thus disregarding also essential aspects of the compresence of spectral and spatial features of the investigated structures.

Here, we present via near-field hyper-spectral imaging (HSI), a detailed sub-wavelength optical characterization of the multipolar resonances formed in a monocrystalline, silicon island, providing simultaneously spectral and spatial information on several resonances in a single measurement. Experimental results are explained and validated by FDTD simulations [18]. Exploiting the etalon effect springing from thick $SiO_2$ layer between the antenna and the bulk Si substrate underneath, we detect the splitting of one resonance into sharper peaks, unrevealing the difference and intimate features of their near-field details. We thus achieve a complete understanding of the effect of the substrate on the resonant modes, filling a gap in a research field that so far has considered the presence of the substrate mostly as a strong limitation of the Mie resonators properties. We also show that intensity and spatial light localization in the near field, strongly depend on the illumination angle, clarifying the crucial role of the excitation configuration that can be exploited as a spatial tuning tool to achieve the spatial overlap between a single emitter and the electric field intensity maximum.

1. Experiment

The investigated samples consist in monocrystalline Si-based islands fabricated via low-resolution optical lithography and plasma etching followed by solid state dewetting [19-22]. A 2 μm thick layer of $SiO_2$ separates the islands from an underlying bulk Si substrate. Details on the fabrication process can be found in [20-22]. In fig. 1a we report a top view Scanning Electron Microscopy (SEM) image of the sample showing several dewetted islands; the details of a single island with a diameter of the order of 300 nm can be observed in the SEM image of the inset of fig.1a.

A scanning near-field optical microscope (SNOM) in collection geometry was used in order to detect the near-field scattered emission of a single island under a tilted illumination configuration (as shown in the sketch of fig. 1b). The spatially-resolved optical maps were recorded by scanning the probe dielectric tip over the sample at a fixed distance (few tens of nm). The illumination source, tilted of an angle $\theta$=30° with respect to the perpendicular axis of the scatterer, is a Supercontinuum laser, whose scattering spectrum, collected with the cooled Si-based CCD camera used in the experiment, is reported in fig. 1c. The signal is fed in a 30cm focal length spectrometer and detected by the CCD camera. At every tip position, the entire spectrum of the sample is collected with a spectral resolution of 0.1 nm. The whole technique allows to perform HSI by collecting a full near-field spectrum at every tip position of the spatial map.

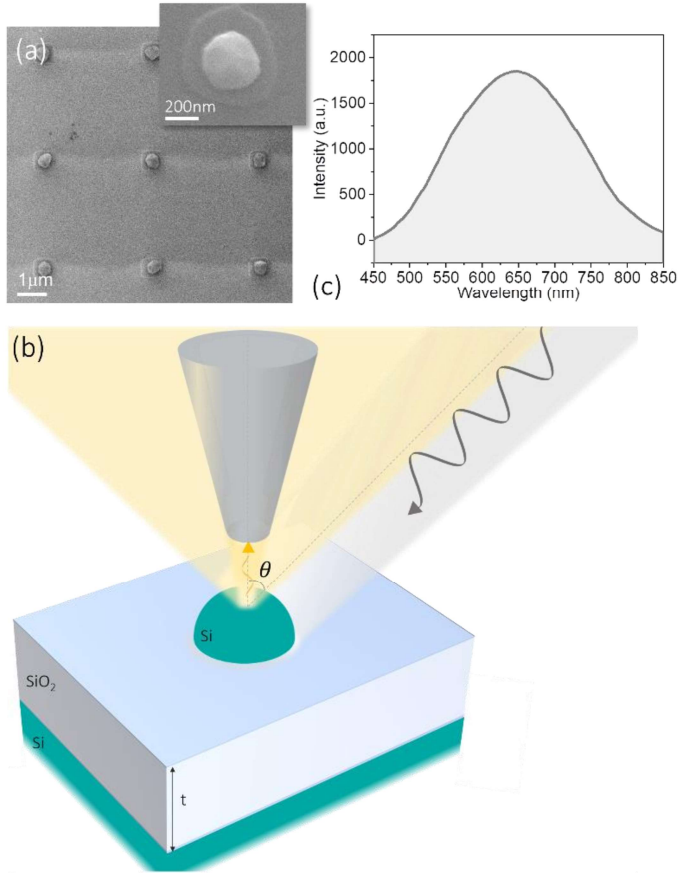

**Figure 1**: **(a)** Top view SEM image of the sample, in which several dewetted islands are shown. The inset displays the magnified SEM image of a single island. **(b)** Sketch of the experimental set up: the tilted illumination (of an angle $\theta$ with respect to the axis perpendicular to the sample) from the Supercontinuum is scattered by the island. The yellow cone is a representation of the angular pattern of scattering. The scattered light is collected by the near-field tip at a distance of $\approx$10nm. **(c)** Emission spectrum of the Supercontinuum source detected by a Si-based CCD camera.

2. Theoretical simulations: Modes evolution with substrate and angle of illumination

We study the modification on the scattering spectra of a single Si-based scatterer as a function of the thickness t of the $SiO_2$ substrate and of the illumination angle $\theta$ by means of FDTD simulations with a commercial software [18]. We consider a hemispherical Si island of diameter d=330nm (fig.2a), illuminated by a Total Field Scattered Field (TFSF) source, which launches a broad-band ($\lambda$=400-1200nm) plane wave from the top, and filters out all the light that has not been scattered. The simulated source is polarized along the x axis, and can be tilted by an angle $\theta$ with respect to the z axis. Power transmission monitors are positioned around the TFSF source in order to obtain the total scattering cross section, and two field monitors intersect the particle (along the xz and yz plane) to monitor the local field intensity. An additional monitor is placed on top of the island in the xy plane at a distance of 10nm from the apex in order to simulate the near-field spatial profile of the electric field intensity. Perfectly Matching Layers (PMLs) are used to prevent unphysical scattering from the simulation boundaries and to mimic semi-infinite substrates. Optical constants for Si and SiO$_2$ are taken from Palik [23].

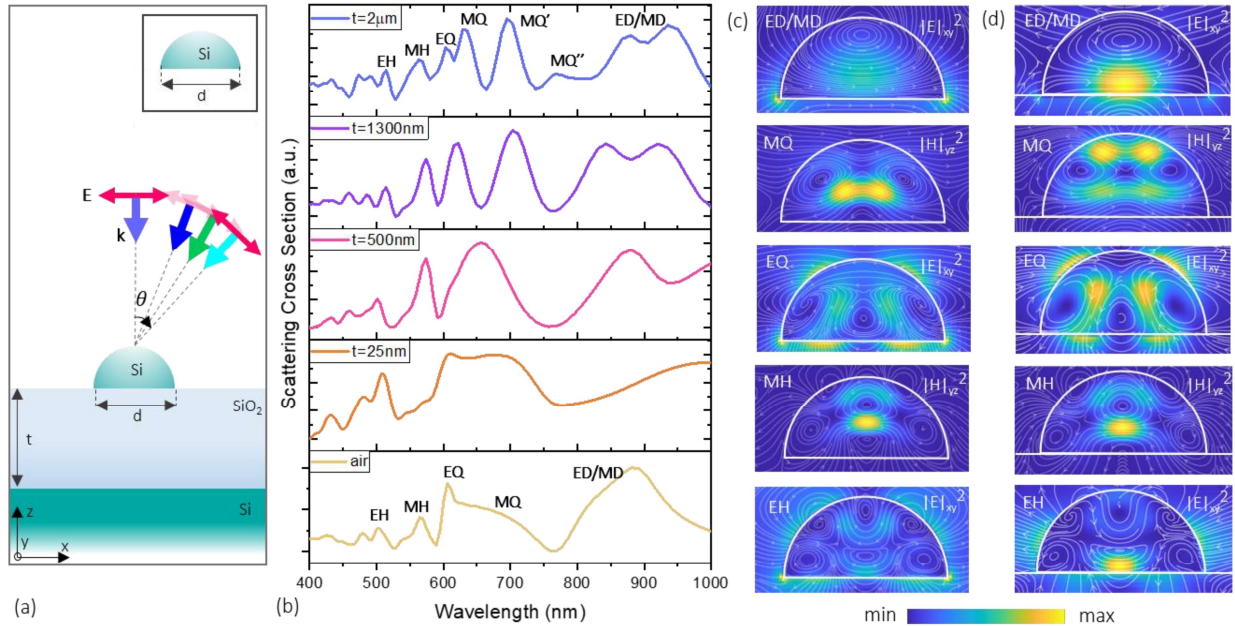

**Figure 2:** **(a)** Sketch of the model used in the FDTD simulations; a Si hemisphere of diameter d is positioned atop a SiO$_2$ layer of thickness t, and a bulk Si substrate. The scatterer is illuminated by an x polarized TFSF source that can be tilted by an angle theta with respect to the z axis. The inset shows the bare case of the island in air. **(b)** FDTD scattering cross sections (from bottom to top) for increasing values of t, t=0 (island in air), t=25nm, 500nm and 1300nm respectively in yellow, orange, magenta and purple. **(c)** Vertical crosscuts of the electromagnetic field intensity and current loops (represented by white arrows) of the multipolar resonances of the hemisphere in air. The maps, acquired at the wavelengths of the yellow spectrum, are reported from top to bottom in order of increasing energy: ED/MD ($\lambda$=800nm), MQ ($\lambda$=660nm), EQ ($\lambda$=605nm), MH ($\lambda$=570nm) and EH ($\lambda$=504nm). **(d)** Vertical crosscuts of the electromagnetic field intensity and current loops (represented by white arrows) of the multipolar resonances of the hemisphere on a SiO$_2$ layer of thickness t=2$\mu$m. The maps, acquired at the wavelengths of the blue spectrum, are reported from top to bottom in order of increasing energy: ED/MD ($\lambda$=770nm), MQ ($\lambda$=628nm), EQ ($\lambda$=603nm), MH ($\lambda$=563nm) and EH ($\lambda$=514nm).

Fig. 2b reports the total scattering cross sections of the Si hemisphere, illuminated under normal incidence, for different values of the substrate thickness t. In particular, it is possible to follow the evolution of the modes from the case of the island in air (yellow spectrum), to the configuration with the maximal SiO$_2$ thickness of 2$\mu$m (blue spectrum). The latter is the actual value of t in the experimentally investigated sample. The scattering cross sections for three selected values of t (i.e. t=25nm, 500nm and 1300nm), are reported in Fig. 2c, in orange, magenta and purple, respectively. The identification of these modes (electric hexapole EH, magnetic hexapole MH, electric quadrupole EQ, magnetic quadrupole (MQ) and electric and magnetic dipole ED/MD) is based on their field intensity profile inside the particle and the corresponding current loops (fig. 2c).

First, electric and magnetic field intensity inside the hemisphere suspended in air allow to identify the multipolar resonances, from the fundamental to higher-energy ones. In the near-field, the electric (magnetic) field profiles in the xz (yz) plane for electric (magnetic) resonances show bright lobes inside, corresponding to the poles generated by current loops (white arrows in Fig. 2 c). Effects of symmetry breaking with respect to the ideal case of a sphere in air (in which ED and MD are easily distinguishable) must be considered [24, 25] and are beyond the scope of this paper. We therefore refer to the broad band centered around 900 nm as ED/MD. We also identify the main higher-order resonances as MQ, EQ, MH and EH, of which we report the vertical crosscuts of E or H respectively at $\lambda$ = 660 nm, 605 nm, 570 nm and 504 nm.

The same reasoning was adapted to assign the resonances of the particle sitting atop a $SiO_2$ substrate while t increases. It was recently shown ([8, 20]) that by increasing t, constructive and destructive interference of the incident light reflected from the $SiO_2$/Si-bulk interface arises, resulting into strong variations in the driving field of the resonator. The Mie scattering efficiently out-couples the light interfering in the etalon and re-directs light at smaller angles with respect to the incident beam. From this combination structural colors spring, covering the full visible spectrum with resonances with a high intensity contrast between maxima and minima. However, a detailed identification of the actual physical origin of these sharp peaks has never been attempted so far. While increasing t, we can follow the evolution of the vertical profiles of the electric (magnetic) field intensity along the xz (yz) plane of the main electric (magnetic) resonances in the scatterer (fig.2d for the main resonances of the blue spectrum on top of fig. 2b, corresponding to the nominal $SiO_2$ thickness t=2μm). Specifically, the maps of ED/MD, MQ, EQ, MH and EH correspond to $\lambda$=770nm, 628nm, 603nm, 563nm and 514nm, respectively. The original broad band corresponding to MQ in the yellow spectrum (scatterer in air) splits into three sharp peaks, all identifiable as magnetic quadrupole resonances from the analyzed current loops, to which we refer as MQ, MQ' and MQ'' (more details are given in Supplementary Information, Fig. S1). The coupling between the scatterer and the interference arising from the $SiO_2$ layer, combined with the analysis of crosscuts of the electric/magnetic field for every peak, therefore allows to identify the modes and obtain information about the multipole characteristics of the island spectrum without any needs of multipolar decomposition [15].

An analogue method of identification is used in order to monitor the changes in the scattering properties of the island with respect to the angle of illumination (fig. 3a). We compute the FDTD scattering cross sections of a Si hemisphere for t=2μm layer of $SiO_2$ and Si bulk for different angles of illumination $\theta$: the spectra are displayed in blue for $\theta$=0°, in dark blue for $\theta$=20°, in green for $\theta$=30° and in cyan for $\theta$=40°. The peaks arising from the coupling of the Mie modes with the etalon, are sensible to the angle of illumination, differently with respect to the scattering cross section of the island on a thin substrate, that results practically unchanged for different $\theta$ (as shown in Fig. S2 of Supplementary Information). In fact the broad MH and MQ peaks split due to interference with the etaloning resonances, in which the angle of incidence plays a fundamental role. Specifically, for $\theta$=30°, a second MH appears around $\lambda$=576nm (that we call MH'), and one of the three MQs is suppressed, leaving only MQ and MQ' at $\lambda$=657nm and $\lambda$=725nm.

We now focus on the scattering in near-field; in fig. 3b we report in dark green the FDTD near-field spectrum acquired at the center of the island for $\theta$=30°, and the calculated scattering cross section for the same angle of illumination. The two spectra display almost the same resonances but with different relative weights of the peaks; in the near-field, the peaks MH' and EQ broaden and as a consequence they tend to overlap. Provided the high sensitivity of the Mie resonances to large values of t, we analyze the near-field spatial profile of the electric field intensity, achieving a full comprehension of the scatterer's responses in such a specific configuration.

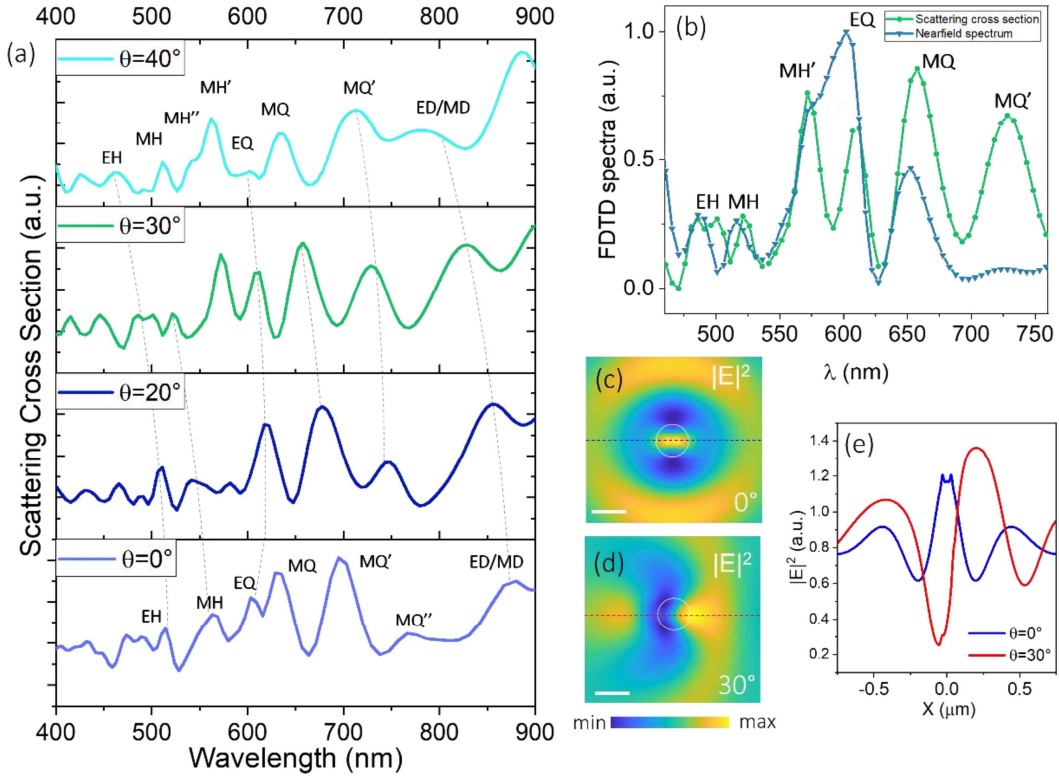

Figure 3: (a) FDTD scattering cross sections for different angles of illumination, from bottom to top: $\theta$=0° in blue, $\theta$=20° in dark blue, $\theta$=30° in green and $\theta$=40° in cyan. (b) FDTD scattering cross section in green and FDTD near-field spectrum, in dark green, acquired on the center of a field monitor at a distance of 10nm from the hemisphere surface for $\theta$=30°. The peaks are labelled according to the identification of the multipolar resonances under tilted illumination EH, MH, MH', EQ, MQ and MQ'. (c), (d): Near-field FDTD spatial distribution of the electric field intensity of MQ for $\theta$=0° and $\theta$=30°. The white scale bar corresponds to 200nm. (e) Horizontal cuts of the near-field|E|² along the dashed blue (red) dashed lines displayed in (c) and (d).

The near-field distributions of the resonances are strongly affected by the illumination direction. In fig. 3c and d are shown the near-field FDTD maps of the electric field intensity of the MQ' mode generated under normal incidence (0°) and tilted incidence (30°), respectively. In the latter case, the electric field intensity is not symmetrically distributed around the scatterer, and we clearly observe that the hotspot changes its shape and spatially shifts to the right. The shift is of about 250 nm, as highlighted by fig. 3e, where the horizontal cross cuts through the center of the maps are compared (respectively in blue and red). These cross cuts show that also the maximum electric field intensity is influenced by the illumination angle. Fig. 3 c-e shows that by changing the illumination angle by 30° on the left half of the Si-island the signal intensity drops from bright to dark, while in the region around the right border of the island we observe a net increase of the electric field intensity.

3. Near-field Hyper Spectral Imaging Experiment

In the following, we provide a detailed near-field HSI analysis of the higher order multipolar modes in a single all-dielectric nanoparticle. Fig. 4 summarizes the results of a 1.6 μm x 1.2 μm SNOM scan on a Si dewetted island (fig. 1a) of diameter d=330nm, under the illumination of a super-continuum source tilted by 30°. The scan step is 50nm/pixel, and the sub-wavelength spatial resolution of about 200 nm is related with the dimensions of the tip. The normalized near-field spectrum (in purple) collected at the apex of the island is reported in fig. 4a, together with the FDTD near-field spectrum acquired at the center of the near-field monitor (in dark green) (details on the normalization of the spectrum are reported in Figure S3 of Supplementary Information). Theoretical and experimental spectra are in good agreement (Fig. 4a), allowing to identify, in order of decreasing energy, the main high-order multipolar resonances that result from the resonant modes of the antenna with the etalon effect, from EH to MQ'. In agreement with the previous assessment of scattering cross-section and near-field spectrum for $\theta$ = 30° shown in FDTD simulations where the two peaks MH' and EQ spectrally overlap (Fig. 3b), the experimental SNOM near-field spectrum at those wavelengths displays a single peak, dominated by the contribution of the EQ resonance (Fig. 4a). Exploiting our HSI technique, we provide the experimental near-field spatial profiles represented in the maps of fig. 4b, that are obtained by plotting the intensity integrated around the main peaks in the purple spectrum of Fig. 4a ($\lambda$ = 490 nm (EH), $\lambda$= 530 nm (MH), $\lambda$ = 570 nm (EQ), $\lambda$ = 645 nm (MQ) and $\lambda$ = 735 nm (MQ')).

During a SNOM scan, scattering spectra and morphology of the sample are simultaneously acquired. Thus, we can overlap optical and morphological maps (the island is highlighted by white circles in Fig. 4b) and directly compare them with simulations. Experimental and simulated maps show striking similarities in the spatial symmetry of the resonances that are typical of each mode and in the relative weights of the lobes of each mode, indicating preferential directions of scattering (Fig. 4b and 4c). The asymmetric "ring crowns" are well reproduced, both in intensity, spatial location and distance from the island. Under tilted illumination, higher energy modes, like EH and MH, display a near-field scattered intensity that is unbalanced in opposite direction with respect to lower energy modes, like MQ and MQ' (towards left for EH and MH, towards right for MQ, MQ'). The EQ mode instead, exhibits a more symmetric distribution both in theory and experiment. The symmetry of EQ is well recognizable in the experimental map, where a vertical dip in the intensity of the signal is visible. In particular, the signal intensity of the MQ' mode reproduces with high fidelity the expected spatial modulation characterized by a bright lobe on the left side of the island and a clear signal dip on the opposite side. Deviations between theory and experiment can be ascribed to several factors ascribed to shape and composition differences [26, 27].

These results show that, in the perspective of maximizing the surface enhancement in absorption or in Raman scattering of a single molecule or other quantum system [3], it is possible to control the spatial overlap between the quantum center and the electric field hot spot by acting on the illumination angle. This represents an interesting alternative way to the most common approach based on the accurate positioning at the fixed field maximum.

In conclusion, we provided a complete near-field hyper-spectral imaging study of the modes arising from the combination of the Mie resonances of an all-dielectric dewetted island with the etalon effect due to a thick $SiO_2$ layer below the scatterer. By means of FDTD simulations we achieved a full comprehension of the sensitivity of magnetic light with respect to the thickness of the substrate and the angle of illumination. The combination of numerical and experimental results offers a detailed analysis of how such sensitivity affects the near-field scattering pattern of the antenna, suggesting a novel approach to control the spatial overlap of a single emitter with localized electric field maxima: instead of placing the emitter on the hot spot, the hot spot can be moved on the emitter position offering a versatile tuning tool to adjust the mutual coupling.

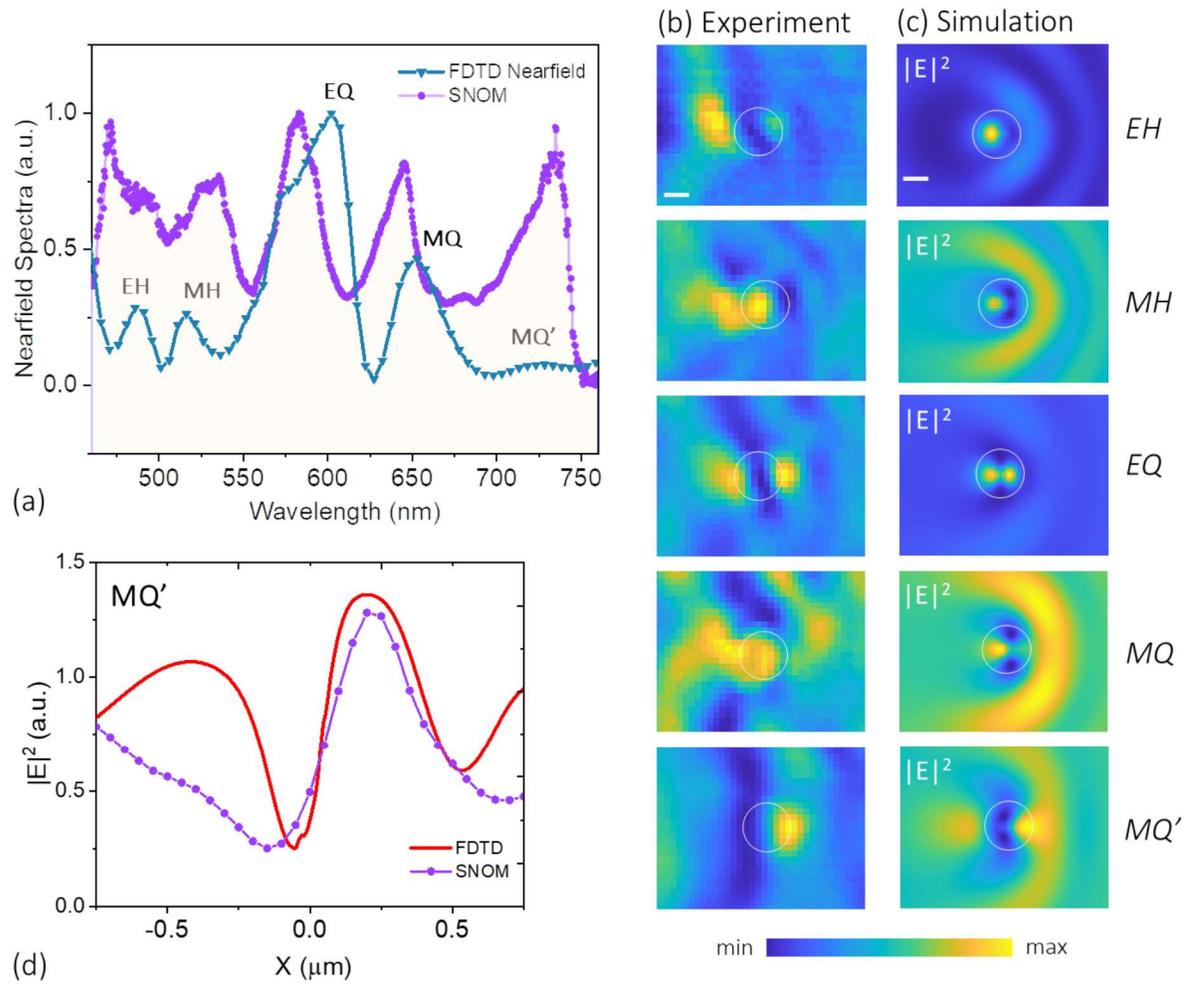

**Figure 4: (a)** Normalized SNOM scattering spectrum collected on top of the dewetted island of d=330nm (purple dots) and FDTD near-field spectrum acquired at 10nm distance from the top center of a hemisphere of the same diameter (in dark green). **(b)** Experimental SNOM maps filtered around the central wavelength of the purple spectrum in (a), respectively relative to the EH ($\lambda$=490nm), MH ($\lambda$=530nm), EQ ($\lambda$=570nm), MQ ($\lambda$=645nm) and MQ' $\lambda$=735nm modes. **(c)** FDTD maps of the electric field intensity of the multipolar resonances detected in the dark-green spectrum: EH ($\lambda$=485nm), MH ($\lambda$=521nm), EQ ($\lambda$=602nm), MQ ($\lambda$=654nm) and MQ' $\lambda$=728nm. All white scalebars correspond to 200nm. **(d)** Horizontal cuts along the central part of the experimental and the FDTD maps relative to the MQ' resonance.

References

[1] Kuznetsov, A. I.; Miroshnichenko, A. E.; Brongersma, M. L.; Kivshar, Y. S.; Luk'yanchuk, B. Optically resonant dielectric nanostructures. *Science* **2016**, 354, 6314.

[2] Hancu, I. M.; Curto, A. G.; Castro-Lopez, M.; Kuttge, M.; van' Hulst, N. F. Multipolar Interference for Directed Light Emission. *Nano Lett*. **2014**, 14, 166–171.

[3] Suárez, I.; Wood, T.; Martinez Pastor, J. P.; Balestri, D.; Checcucci, S.; David, L.; Favre, L.; Claude, J.B.; Grosso, D; Gualdrón-Reyes, A. F.; Mora-Seró, I.; Abbarchi, M. and M. Gurioli. Enhanced nanoscopy of individual CsPbBr3 perovskite nanocrystals using dielectric sub-micrometric antennas. *APL Materials* **2020**, 8, 021109.

[4] Dotti, N.; Sarti, F.; Bietti, S.; Azarov, A.; Kuznetsov, A.; Biccari, F.; Vinattieri, A.; Sanguinetti, S.; Abbarchi, M. and Gurioli, M. Germanium-based quantum emitters towards a time-reordering entanglement scheme with degenerate exciton and biexciton states. *Phys. Rev. B* **2015**, 91(20), 205316.

[5] Capretti, A.; Lesage, A. and Gregorkiewicz, T. Integrating Quantum Dots and Dielectric Mie Resonators: A Hierarchical Metamaterial Inheriting the Best of Both. *ACS Photonics* **2017**, 4, 2187-2196.

[6] Tognazzi, A.; Okhlopkov, K. I.; Zilli, A.; Rocco, D.; Fagiani, L.; Mafakheri, E.; Bollani, M.; Finazzi, M.; Celebrano, M.; Shcherbakov, M. R.; Fedyanin, A. A. and De Angelis, C. Third-harmonic light polarization control in magnetically resonant silicon metasurfaces. *Opt. Express* **2021**, 29, 8, 11605.

[7] Alhalabya, H.; Zaraket, H. and Principe, M. Enhanced Photoluminescence with Dielectric Nanostructures: A review. *Results in Optics* **2021**, 3, 100073.

[8] Todisco, F.; Malureanu, R.; Wolff, C.; Gonçalves, P. A. D.; Roberts, A. S.; Mortensen, N. A. and Tserkezis, C. Magnetic and electric Mie-exciton polaritons in silicon nanodisks. *Nanophotonics* **2020**, 9, 4, 803-814.

[9] Van de Groep, J. and Polman, A. Designing dielectric resonators on substrates: Combining magnetic and electric resonances. *Opt. Express* **2013**, 21(22), 26285–26302.

[10] Hao, F.; Nordlander, P. Efficient dielectric function for FDTD simulation of the optical properties of silver and gold nanoparticles. *Chem. Phys. Lett.* **2007**, 446, 115–118.

[11] Spinelli, P., Verschuuren, M. & Polman, A. Broadband omnidirectional antireflection coating based on subwavelength surface Mie resonators. *Nat Commun* **2012**, 3, 692.

[12] Evlyukhin, A. B.; Novikov, S. M.; Zywietz, U.; Eriksen, R. L.; Reinhardt, C.; Bozhevolnyi, S. I.; Chichkov, B. N. Demonstration of Magnetic Dipole Resonances of Dielectric Nanospheres in the Visible Region. *Nano Lett.* **2012**, 12, 3749–3755.

[13] Kuznetsov, A. I.; Miroshnichenko, A. E.; Fu, Y. H.; Zhang, J.; Luk'yanchuk, B. Magnetic Light. *Sci. Rep*. **2012**, 2, 492–1–492–6.

[14] Coenen, T.; Van de Groep, J. and A. Polman. Resonant modes of single silicon nanocavities excited by electron irradiation. *ACS Nano* **2013** 7(2), 1689–1698.

[15] Habteyes, T.G.; Staude, I.; Chong, K. E.; Dominguez, J.; Decker, M.; Miroshnichenko, A.; Kivshar, Y. and Brener, I. Near-Field Mapping of Optical Modes on All-Dielectric Silicon Nanodisks. *ACS Photonics* **2014**, 1, 794–798.

[16] Miroshnichenko, A., Evlyukhin, A., Yu, Y. et al. Nonradiating anapole modes in dielectric nanoparticles. *Nat Commun* **2015**, 6, 8069.

[17] Bakker, R.M.; Permyakov, D.; Feng Yu, Y.; Markovich, D.; Paniagua-Domínguez, R.; Gonzaga, L.; Samusev, A.; Kivshar, Y., Luk'yanchuk, B. and Kuznetsov, A. I. Magnetic and Electric Hotspots with Silicon Nanodimers. *Nano Lett*. **2015**, 15, 3, 2137–2142.

[18] FDTD Solutions, Lumerical Solutions, Inc., http://www.lumerical.com.

[19] Thompson, C. V. Solid-State Dewetting of Thin Films. *Annu. Rev. Mater. Res*. **2012,** 42(1), 399–434.

[20] Toliopoulos, D.; Khoury, M.; Bouabdellaoui, M.; Granchi, N.; Claude, J.-B.; Benali, A.; Berbezier, I.; Hannani, D.; Ronda, A.; Wenger, J.; Bollani, M.; Gurioli, M.; Sanguinetti, S.; Intonti, F. and Abbarchi, M. Fabrication of spectrally sharp Si-based dielectric resonators: combining etaloning with Mie resonances. *Opt. Express* **2020**, 28, 25, 37734-37742.

[21] Abbarchi, M.; Naffouti, M.; Vial, B.; Benkouider, A.; Lermusiaux, L.; Favre, L.; Ronda, A.; Bidault, S. ; Berbezier, I. and Bonod, N. Wafer scale formation of monocrystalline silicon-based mie resonators via silicon-on-insulator dewetting, *ACS Nano* **2014** 8 (11), 11181–11190.

[22] Naffouti, M.; Backofen, R.; Salvalaglio, M.; Bottein, T.; Lodari, M.; Voigt, A.; David, T.; Benkouider, A. ; Fraj, I.; Favre, L.; Ronda, A.; Berbezier, I.; Grosso, D.; Abbarchi, M. and M. Bollani. Complex dewetting scenarios of ultrathin silicon films for large-scale nanoarchitectures. *Sci. Adv*. **2017**, 3(11), 1472.

[23] Palik, E. D. Handbook of Optical Constants of Solids, Academic, 1985.

[24] Butakov, N. A. & Schuller, J.A. Designing Multipolar Resonances in Dielectric Metamaterials. *Scientific Reports* **2016***,* 6, 38487.

[25] Berzins, J.; Indrisiunas, S.; Van Erve, K.; Nagarajan, A.; Fasold, S.; Steinert, M.; Gerini, G.; Gecys, P.; Pertsch, T. ; Baumer, S. M. B. and Setzpfandty, F. Direct and High-Throughput Fabrication of Mie Resonant Metasurfaces via Single-Pulse Laser Interference. *ACS Nano* **2020**, 14, 5, 6138–6149.

[26] Bollani, M., Salvalaglio, M., Benali, A. et al. Templated dewetting of single-crystal sub-millimeter-long nanowires and on-chip silicon circuits. *Nat Commun* **2019**, 10, 5632.

[27] Leroy, F.; Saito, Y.; Curiotto, S.; Cheynis, F.; Pierre-Louis, O. and Müller, P. Shape transition in nano-pits after solid-phase etching of SiO2 by Si islands. *Appl. Phys. Lett*. **2015**, 106, 191601.

# Supplementary Information

## Figure S1

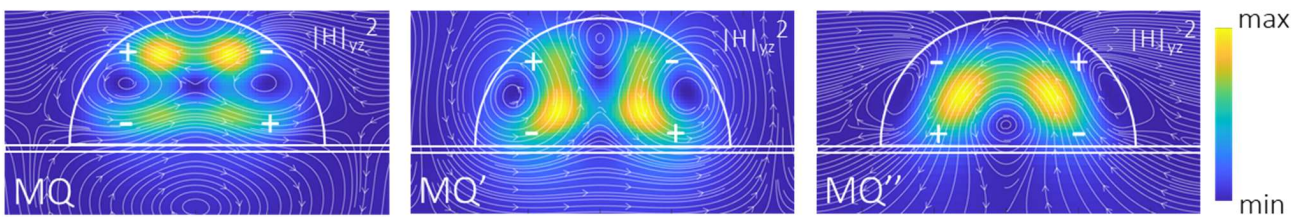

FDTD vertical crosscuts of the magnetic field intensity and current loops (represented by white arrows) of the triple magnetic quadrupole resonances of the hemisphere on a $SiO_2$ layer of thickness t=2mm at normal incidence. Respectively MQ, MQ' and MQ'' at λ=628nm, 695nm and 766nm. The corresponding poles are labeled with + and – signs (white). MQ'' presents poles of inverted signs with respect to MQ and MQ'.

## Figure S2

Here we show that the peaks in the scattering cross section of a Si island on a thin $SiO_2$ substrate (25nm) are not affected by the angle of illumination.

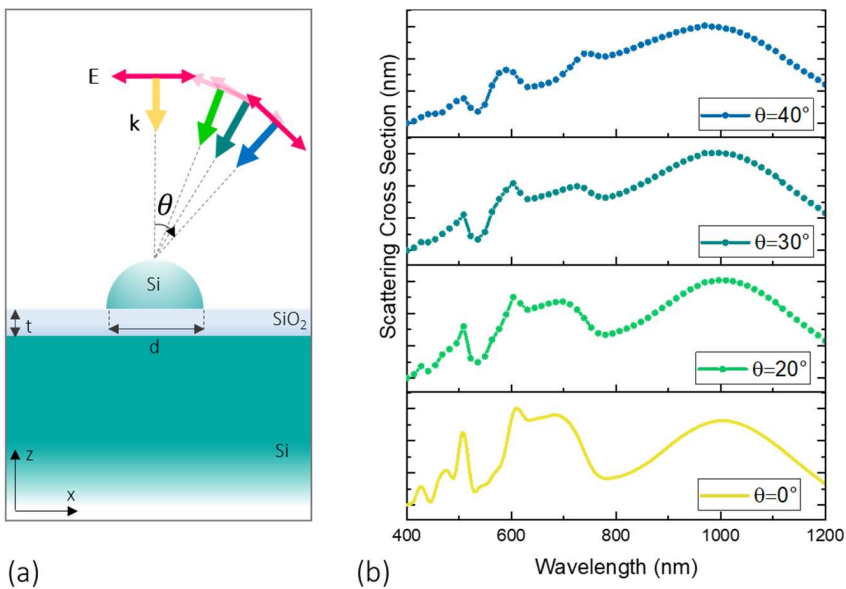

(a) (b)

(a) Sketch of the model used in the FDTD simulations; a Si hemisphere of diameter d is positioned atop a $SiO_2$ layer of thickness t=25nm, and a bulk Si substrate. The scatterer is illuminated by an x polarized TFSF source that can be tilted by an angle theta with respect to the z axis. (b) FDTD scattering cross sections for different angles of illumination, from bottom to top: $\theta$=0° in yellow, $\theta$=20° in light green, $\theta$=30° in dark green and $\theta$=40° in blue.

## Figure S3

This figure represents the method used in order to normalize the SNOM spectra reported in the manuscript. We evaluate the background signal between the dewetted islands by mediating the spectrum in a selected region of the optical map. Then the near-field spectrum on the single island is divided by the mean spectrum reported in Figure S3c.

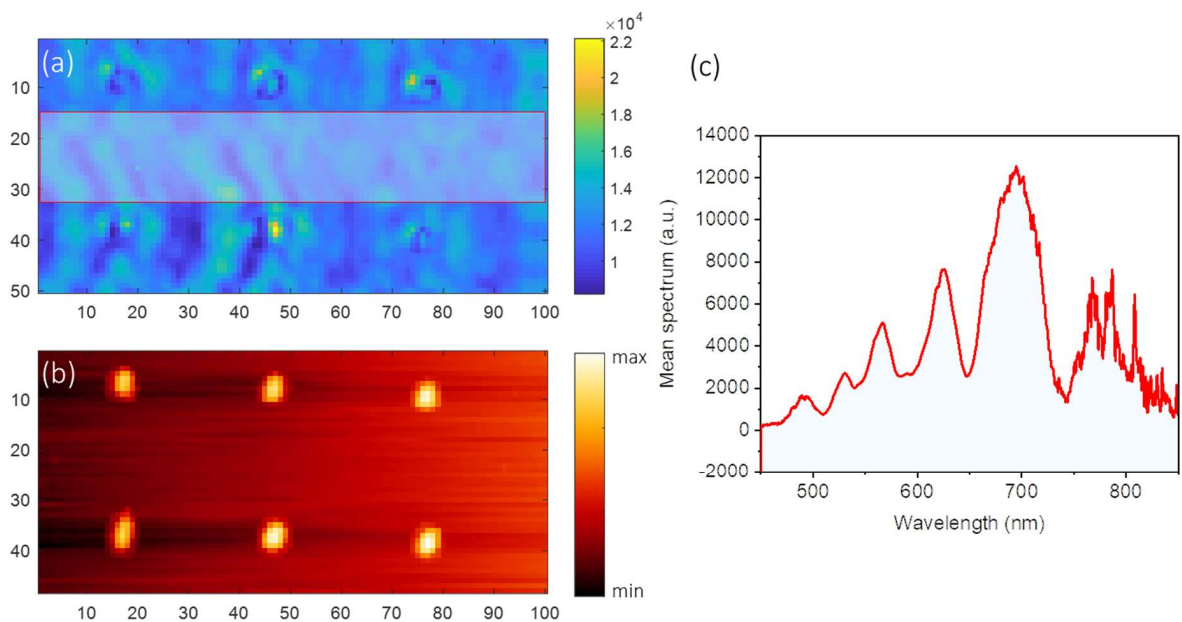

(a) SNOM optical map filtered around a broad wavelength range (400nm-900nm), acquired on a 10 µm x 6 µm region with 100nm spatial step. (b) SNOM topography acquired simultaneously during the scan, in which six ordered dewetted island are visible. (c) Spectrum obtained by mediating in the rectangular area (represented in (a)); the area was chosen by excluding the islands, so that the mean spectrum can be considered as the background signal used to normalize the spectra of the islands.